\documentclass[11pt]{article}

\usepackage{digra}
\usepackage{authblk}
\usepackage{amsmath}       
\usepackage{graphicx}      

\usepackage[table]{xcolor} 
\usepackage{colortbl}      

\usepackage{array}         
\usepackage{tabularx}      
\usepackage{booktabs}      
\usepackage{arydshln}      

\usepackage[authordate,minnames=1,maxnames=2,maxbibnames=10,minbibnames=7]{biblatex-chicago}

\AtEveryBibitem{\clearfield{doi}\clearfield{urlyear}\clearfield{urlmonth}\clearfield{urlday}}
\DeclareFieldFormat[game]{title}{\mkbibemph{#1}}
\DefineBibliographyStrings{english}{%
  references = {},
}

\let\originalcite\cite
\renewcommand*{\cite}[1]{(\originalcite{#1})}

\definecolor{theme1}{RGB}{70, 120, 200}   
\definecolor{theme2}{RGB}{100, 175, 120}  
\definecolor{theme3}{RGB}{225, 170, 80}   
\definecolor{theme4}{RGB}{255, 220, 245}  

\title{How to be Non-Human : A Thematic Analysis of Animal Embodiment in VR Games }

\author[1,*]{Siqi Yu}
\author[2,*]{Shuai Liu}
\author[3]{Yiqing Tian}
\author[4,**]{Mar Canet Sola}

\affil[1]{The Hong Kong University of Science and Technology (Guangzhou), China, yusiqi73@gmail.com}
\affil[2]{Academy of Media Arts Cologne, Germany, shuai.liu@khm.de}
\affil[3]{Goldsmiths, University of London, UK, itian001@gold.ac.uk}
\affil[4]{BFM, Tallinn University, Estonia, mar.canet@gmail.com}
\affil[*]{Equal contribution}
\affil[**]{Corresponding author}

\date{\vspace{-60pt}}
\begin{document}
\pagenumbering{gobble} 
\newpage
\pagenumbering{arabic}  
\addvspace{-1\baselineskip}
   \maketitle
   \addvspace{-1\baselineskip}
    \copyrightnotice

\section*{ABSTRACT}
This study employs a reflexive thematic analysis to systematically examine the design patterns of 48 first-person Virtual reality (VR) animal avatar games. The research identifies four primary design themes: Animal Biomimicry, Limited Animal Simulation, Hybrid Human-Animal Features, and Human Behavior with Animal Avatar. The analysis reveals that approximately 77 percent of the games remain grounded in human-centered interaction logic, with animal forms primarily serving as visual representations. The study highlights the core tension between authenticity and usability in current VR animal avatar design, and points toward design opportunities for achieving more authentic animal avatar's interactive experience through directions such as controller innovation, unconventional body mapping, and dynamic feedback. This research provides a thematic classification framework for understanding the representation of non-human perspectives in VR games.
\section*{Keywords}
Virtual Reality, Non-human Animal Embodiment, Embodied Interaction,  Reflexive Thematic Analysis, Game Design

\section*{INTRODUCTION}
Virtual reality (VR) technology enables humans to experience different bodily forms, among which animal avatars allow for perceiving and moving from the perspective of non-human animals. Taking on the role of an animal in games enhances enjoyment and player experience. Players' experienced autonomy, intuitive control, and oculomotor vary significantly depending on body features and abilities of the animal \cite{krekhov2019beyond}. A wide variety of VR games now offer animal embodiment experiences, ranging from domestic pets to wild animals. 

Existing research on VR avatars has confirmed that users can develop a sense of ownership over virtual bodies with non-human forms \cite{krekhov2018vr} and that animal embodiment experiences can enhance nature connectedness \cite{ahn2016experiencing} and promote pro-environmental behavior \cite{pimentel2025uses}. However, most of these studies focus on specific cases or prototype systems in laboratory settings, lacking a systematic investigation into the actual design practices of animal avatars in commercial VR games.

This gap raises several critical unanswered questions: How exactly are animal embodiment experiences designed and presented in contemporary VR games? To what extent do these games fulfill the promise of becoming an animal? What design strategies are adopted to address the differences between human and animal bodies? More importantly, given the technological reality that VR hardware is fundamentally designed around the human body, how do game designers balance authenticity, playability, and technical feasibility?

These questions not only pertain to game design practice but also touch upon the potential of VR as a perspective-shifting medium. Animal avatar games constitute a special case. Animal avatar games lie at the intersection of experimental and commercial domains, as well as embodied theory and market expectations. Their designs often reveal how radical body transformation is constrained, compromised, or reinvented in practical applications. Therefore, they provide an ideal observational window for examining how VR avatar theory is implemented in industrial contexts.

This study employs reflexive thematic analysis to systematically examine 48 first-person VR animal avatar games, mapping out the patterns and logic of current design practices. We focus on the first-person perspective because it is a necessary condition for inducing the illusion of body ownership \cite{maselli2013building} and represents the most direct attempt to simulate animal subjectivity. The aims of this study are: 1) to extract themes for VR non-human animal avatar design; 2) to reveal dominant patterns and marginal cases in current design practices; 3) to analyze the relationship between technical constraints, design conventions, and experiential goals underlying design choices; and 4) to suggest possible directions for future designs that are more diverse and authentic in representing animal embodiment.

\section*{RELATED WORK}

\subsection*{VR Embodied Interaction and Virtual Body Ownership}

\subsubsection*{Embodied Interaction Theory}

Embodied interaction research emerged in the late 1990s as a response to graphical user interface paradigm. The introduction of the Tangible User Interface (TUI) \autocite{ishii1997tangible} proposed creating seamless interfaces between people, bits and atoms by giving digital information a physical form. Around the same time, the Rubber Hand Illusion (RHI) experiment \autocite{botvinick1998rubber} showed how synchronized visual and tactile stimulation could make people attribute touch sensations to external objects - this laid the groundwork for studying body ownership.

The embodied interaction framework \autocite{dourish2001action} brought phenomenological ideas into human-computer interaction, particularly Heidegger's \textit{Zuhandenheit} (ready-to-hand) - the idea that tools fade into the background when we use them skilfully - and Merleau-Ponty's body phenomenology, which sees the body as our means of experiencing the world rather than just an object. This framework argued that embodiment isn't just about physical properties - it also involves social presence and participatory status, establishing the body as the central site where meaning is created. 

\subsubsection*{Body Ownership in VR}

Advances in neuroimaging technology took body ownership research beyond behaviour and into neural mechanisms. Functional Magnetic Resonance Imaging (fMRI) studies revealed that the rubber hand illusion involves multisensory integration brain regions \autocite{ehrsson2005touching}, whilst the Video ergo sum study \autocite{lenggenhager2007video} extended body ownership illusions from limbs to the full body. Participants viewed themselves from behind through a head-mounted display whilst receiving synchronized tactile stimulation, and reported feeling as though they were located at the virtual body's position.

VR technology provided an ideal platform for studying the Illusion of Virtual Body Ownership (IVBO). Research in immersive VR environments demonstrated that visuotactile synchrony, visuomotor synchrony, and even brain-computer interfaces could all induce virtual arm ownership illusions. VR presence was distinguished into Place Illusion (PI) - the feeling of actually being in the virtual environment - and Plausibility Illusion (Psi) - the feeling that virtual events are really happening \autocite{skarbez2017psychophysical}.

Regarding the impact of virtual avatars on users' behaviors and cognition, the Proteus Effect \autocite{yee2007proteus} revealed that user behaviour is influenced by their avatar's appearance - users with taller avatars were more confident in negotiations, whilst those with more attractive avatars were more friendly in social interactions. This provided a theoretical framework for understanding the psychological effects of virtual embodiment.

\subsubsection*{The Three-Component Framework of Embodiment}

The three-component Sense of Embodiment (SoE) framework \autocite{kilteni2012extending} established working definitions and measurement standards for VR embodiment - including the Sense of Self-Location (feeling located where the virtual body is), Sense of Agency (feeling in control of the virtual body), and Sense of Body Ownership (feeling the virtual body is one's own body). This framework sparked empirical research into body schema plasticity.

The super-long virtual arm experiment demonstrated that people can accept and own limbs that are obviously beyond normal proportions, revealing the adaptive capacity of the body schema. Research into key factors for full-body ownership illusions \autocite{maselli2013building} confirmed that first-person perspective is a necessary condition, whilst realistic humanoid appearance significantly strengthens the illusion. A study on embodying child avatars \autocite{banakou2013illusory} provided supporting evidence
that virtual bodies don't just affect perception - they can reshape cognitive processing. Embodying a child avatar led to overestimation of object sizes and changes in implicit attitudes.

\subsection*{Non-Human Animal Perspectives in Games}

\subsubsection*{Understanding Animal Characters}

Animal characters in video games present diverse forms. A classification based on their relationship with human characters identifies four types - animals as characters, animals as companions, animals as enemies, and animals as environment \autocite{janski2016towards}. Analysis from a philosophical perspective reveals how games either reproduce or challenge anthropocentrism - certain games reinforce human-centred perspectives by treating animals as resources or obstacles, whilst others offer non-anthropocentric experiences by having players embody animal characters \autocite{tyler2022game, chang2019playing}. 

Specific game cases demonstrate different ways of constructing animal agency. Animal Mayhem conveys animal perspectives through third-person viewpoints and destructive gameplay \autocite{caracciolo2021animal}, whilst Stray uses spatial narrative to let players understand the environment through feline embodiment \autocite{khalili2024architecture}. Attitude change research validates the extension of the Proteus Effect to animal avatars - players' implicit attitudes towards animals significantly improved after embodying animal characters \autocite{emmerich2024playing}. The academic community's sustained attention to this topic is reflected in dedicated workshop organisation, which continues to explore how games shape relationships between humans and non-human animals.

\subsubsection*{Educational Value and Empathy Mechanisms}

Educational video games demonstrate the learning value of animal perspectives. A farm animal welfare education game for children significantly increased animal welfare knowledge and beliefs about animals' emotional capacities \autocite{hawkins2019development}. The digital simulated ecosystem Eco promotes environmental awareness by visualising the impact of player actions on the environment \autocite{fjaellingsdal2019gaming}. The teaching value of immersive games for species identification has been validated - players learn animal behaviour and interspecies interactions through Red Dead Redemption 2 \autocite{crowley2021educational}, whilst 76\% of Animal Crossing players learned to identify organisms through the game \autocite{coroller2023video}. 

Cross-media research reveals common principles in animal character design. The influence of virtual animal appearance on empathy shows interaction effects - artificial versus natural body appearance and facial expressions jointly shape user experience \autocite{sierra2020influence}. First-person narratives from animal perspective chatbots significantly improve user empathy towards animals \autocite{li2025animal}. These findings provide a theoretical foundation for animal character design in games.

\subsection*{Embodiment of non-human animals in VR}

The applicability of the Illusion of Virtual Body Ownership (IVBO) to animal avatars was verified for the first time. Research in VR environments demonstrated that virtual body ownership can extend to non-humanoid forms such as tigers, bats and spiders \autocite{krekhov2018vr}. Studies on scorpion, rhinoceros and bird avatars found a positive correlation between IVBO and game enjoyment, with additional body parts (wings, rhino horns, scorpion tails) significantly enhancing the experience \autocite{krekhov2019beyond}. Animal body ownership illusions can, in some cases, surpass those of humanoid avatars. Interaction design strategies encompass third-person companion mode, first-person full-body tracking, and inverse kinematics-based body mapping. The influence of animal avatars on player attitudes and behaviour extends the boundaries of the Proteus Effect's application \autocite{emmerich2024playing}.

The effect of animal embodiment in promoting nature connectedness has gained empirical support. An experiment had participants embody cows on all fours (using cattle prod haptic feedback to induce body transfer) and coral in acidifying oceans, finding that animal embodiment enhanced nature connectedness, with effects lasting a week \autocite{ahn2016experiencing}. In the field of conservation psychology, the development of the Project SHELL system found that embodying loggerhead sea turtles (\textit{Caretta caretta}) can reverse "compassion fatigue" and increase charitable donations. The system uses Oculus Quest and SUBPAC haptic backpack to achieve a multisensory experience \autocite{pimentel2022effects}. Field research at beach music festivals demonstrated that both AR and VR animal embodiment produce strong body transfer and promote pro-environmental behavioural intentions \autocite{pimentel2025uses}. 

Embodying different animal forms presents unique technical challenges and body ownership characteristics. Research on bird flight experiences found that visual-motor feedback is more effective than visual-tactile feedback in inducing a sense of wing ownership \autocite{egeberg2016extending}. Four-legged animal motion mapping achieved a technical breakthrough, with neural network-driven dog avatar systems enhancing the sense of body ownership \autocite{egan2023neurodog}. Research on controlling additional arthropod appendages found that bat wings and scorpion tails, perceived as body extensions, received positive evaluations. Comparative experiments with beaver avatars demonstrated that natural animal bodies produce stronger immersion than robotic bodies \autocite{sierra2020influence}.

\subsection*{Research Questions}

Existing research encompasses multidimensional advances in non-human animal embodiment studies, from animal agency design in video games and mechanised role-playing in board games to non-human animal embodiment effects in VR environments, demonstrating cross-media practices of non-human animal perspectives. However, the interaction design patterns of the numerous VR non-human animal role-playing games currently on the market have not yet been systematically reviewed or critically analysed.

This study employs thematic analysis to systematically examine non-human animal role-playing games on existing VR platforms, proposing the following questions:

\textbf{RQ1:} How do existing VR games design movement patterns for first-person non-human animal role-play?

\textbf{RQ2:} What experiential outcomes do these designs produce, and what flaws and limitations of VR game design do they expose?

\textbf{RQ3:} Based on the thematic analysis results, how should future VR non-human animal embodied interaction design develop?

\section*{METHOD}
This study employs the Reflexive Thematic Analysis method \cite{braun2019reflecting,braun2023toward} proposed by Braun and Clarke to understand presentation patterns and design logic of being non-human animals in VR games. The analysis process follows six phases, as reflexive thematic analysis emphasizes, the entire process is highly iterative, overlapping, and continuously as the research team's understanding evolves. This section describes the finalized research process according to the six phases and emphasizes the researchers' reflexivity and argumentative pathways.

\subsubsection*{Dataset collection}

The research began by constructing a corpus covering ``first-person non-human animal VR games." All three researchers are long-term VR players and game researchers, and this ``well-played" background profoundly influenced how we identified and understood the games.We first conducted systematic searches on the two major mainstream platforms: Meta Store and Steam.Two researchers examined games sequentially in A–Z order, while one researcher conducted supplementary searches using animal-related keywords. This process identified 76 candidate VR games.

Subsequently, three researchers downloaded and played all games using Meta Quest 3. During gameplay, we systematically played through the VR games in our dataset. Due to the inherent difficulties of taking written notes while immersed in VR environments, we adopted a verbal dictation method during gameplay. Researchers verbally recorded their observations regarding game mechanics, operational experiences, animal embodiment methods, and subjective feelings in real-time. Upon completing gameplay sessions for all titles, these video recordings were transcribed and organized into a structured dataset. To supplement our firsthand experiences and capture additional gameplay paths and edge cases, we also analyzed player gameplay videos on platforms such as YouTube, applying the same verbal recording and transcription process.

\subsubsection*{Phase 1: Familiarising Yourself with the Dataset}

 Through combination of direct gameplay experience, video analysis, and subsequent data organization, researchers gradually familiarized themselves with the dataset and established a comprehensive understanding of the non-human animal experience phenomenon.

\subsubsection*{Phase 2: Coding}

After familiarizing ourselves with the data, the three researchers independently coded all gameplay notes. The coding at this stage was inductive and developed around the following aspects:

\begin{itemize}
    \item Whether the game truly makes players ``become that animal" (degree of first-person embodiment)
    \item Whether movement representation follows human templates, animal templates, or a mixture of both
    \item Whether animal body structure affects interaction methods
    \item How in-game control methods, feedback patterns, and bodily capabilities are presented
    \item Researchers' feelings of inconsistency,incoherence,anthropomorphism during the experience
\end{itemize}

At this stage, coding aimed to capture as much phenomenal potential as possible without presupposing thematic structures. Starting from researchers' experiences, we tagged all features potentially related to ``non-human animal experience." The final coding results were integrated back into the Miro board and served as the foundation for subsequent discussions.

\subsubsection*{Phase 3: Generating Initial Themes}

The research team compiled all codes in regular meetings and began searching for connections between them. We compared how different animal games shaped bodily capabilities, movement logic, and expectations for players' ``becoming-animal." Initial themes emerged at this stage, including:

\begin{itemize}
    \item Degree of anthropomorphism in animal movements
    \item Whether control methods align with animal body structure
    \item Weakening or enhancement of animal abilities
    \item Whether players are required to operate animal bodies ``in human ways"
\end{itemize}

This phase was also the most active part of reflexivity.The team continuously reflected on how our dual perspective as ``human players" and ``researchers" influenced understanding, attempting to identify which themes derived from embodied experience and which originated from design conventions. Researchers also noted the phenomenon of disproportionately high representation of primate (monkey/gorilla) games and treated this as an analytical issue requiring further discussion.

\subsubsection*{Phase 4: Developing \& Reviewing Themes}

At this stage, we repeatedly examined the boundaries, conceptual implications, and degree of data support for candidate themes. We particularly addressed an important analytical issue: should primates (monkey, gorilla) be analyzed using the same framework as other animals?

Primates accounted for a large sample size, and their high similarity to humans in body structure, hand capabilities, and movement logic made them prone to overshadowing other animal types in coding and theme generation. This phenomenon compelled the team to engage in deep reflection: if all animals are treated equally, would the analysis unfairly lean toward anthropomorphized animals rather than animalized experiences?

Therefore, we conducted more detailed internal divisions of primate games, and after confirming their internal heterogeneity, reintegrated them into higher-level thematic structures to ensure the thematic framework had explanatory power for different animal types.

Meanwhile, the research team removed 28 games that did not meet the criteria (e.g. non-first-person, non-animal role-playing, or duplicate series entries), ultimately forming a valid dataset of 48 games.

\subsubsection*{Phase 5: Refining, Defining, \& Naming Themes}

As the analysis progressed, we gradually refined higher-level thematic structures capable of explaining most games. At this stage, the team constantly returned to coding, gameplay notes, and screenshots to ensure each theme had data support and clearly defined conceptual boundaries.

We ultimately determined four themes (detailed in the chapter  ``Themes Description"), which characterize the main patterns of ``non-human animal perspective" design in contemporary VR games from four dimensions: animal body structure, interaction methods, movement generation logic, and degree of anthropomorphism. These themes represent the conceptual understanding continually revised by researchers through reflection.

\subsubsection*{Phase 6: Writing Up}

In the writing phase, the research team integrated the four major themes into the main analysis section of the paper. During the same time, the team further reflected on how our roles ,such as experienced players, VR researchers, and embodied experiencers,influenced the formation of themes. For example, the high proportion of primate games, game design's long-term dependence on human controller operation logic, and researchers' own bodily expectations may all have shaped how we understood animality.

Therefore, the writing phase is not only a presentation of results, but also a transparent presentation of the analysis process, enabling readers to understand the path of theme generation, the researchers' positionality behind it, and how the analysis was continuously reshaped through discussion.

\section*{THEME DESCRIPTION}

\begin{table}[htbp]
    \centering
    \caption{Themes and Coding Definitions}
    \label{tab:themes_final}
    \begin{tabularx}{\textwidth}{@{}
        >{\hsize=0.6\hsize\raggedright\arraybackslash}X
        >{\hsize=1.4\hsize\raggedright\arraybackslash}X
        @{}}
        \toprule
        \rowcolor{gray!20}
        \textbf{Themes Name} & \textbf{Coding} \\
        \midrule
        \rowcolor{theme1}
        \textbf{Animal Biomimicry} &
        Specific Movement Pattern Simulation \newline Coordinated Movement Beyond Mapped Points \\
        \hdashline
        \rowcolor{theme2}
        \textbf{Limited Animal Simulation} &
        Animal Perspective, Body Shape, and Scale Reproduction \newline Hardware-Constrained Movement \\
        \hdashline
        \rowcolor{theme3}
        \textbf{Hybrid Human-Animal Features} &
        Interactions Incorporating Animal Ecological Features \newline Anthropomorphized Tasks \\
        \hdashline
        \rowcolor{theme4}
        \textbf{Human Behavior with Animal Avatar} &
        Exclusively Human Movement and Interaction Logic \newline Human Body as Foundation \\
        \bottomrule
    \end{tabularx}
\end{table}

\begin{table}[htbp]
    \centering
    \caption{Game Classification by Themes}
    \label{tab:game_classification}
    \begin{tabularx}{\textwidth}{@{}
        >{\hsize=0.8\hsize\raggedright\arraybackslash}X
        >{\hsize=1.2\hsize\raggedright\arraybackslash}X
        @{}}
        \toprule
        \rowcolor{gray!20}
        \textbf{Theme} & \textbf{Games (n=48)} \\
        \midrule
        \rowcolor{theme1}
        \textbf{Theme 1: Animal Biomimicry} \newline (n=2) &
        \textit{VR Pigeons}, \textit{Squid Smack} \\
        \hdashline
        \rowcolor{theme2}
        \textbf{Theme 2: Limited Animal Simulation} \newline (n=9) &
        \textit{I am Cat}, \textit{Catlateral Damage}, \textit{No More Rainbows Review}, \textit{Jupiter \& Mars}, \textit{I am Hamster}, \textit{Zombie Ants VR}, \textit{A Cats Life}, \textit{Monket Doo}, \textit{Gorilla Tower Jump} \\
        \hdashline
        \rowcolor{theme3}
        \textbf{Theme 3: Hybrid Human-Animal Features} \newline (n=12) &
        \textit{Eagle Flight}, \textit{Flying Squirrel Chase}, \textit{I am Monkey}, \textit{Wreckin' Raccoon}, \textit{Dog Plan}, \textit{Tentacular}, \textit{Catana}, \textit{Me, You and Kaiju}, \textit{Monster: Titan's Playground}, \textit{I am Monket Tower Defense},
        \textit{Gorilla vs 100 Men},
        \textit{Gorilla Attack}\\
        \hdashline
        \rowcolor{theme4}
        \textbf{Theme 4: Human Behavior with Animal Avatar} \newline (n=25) &
        \textit{Animal Blocks}, \textit{Animal Company}, \textit{Grand Theft Animals}, \textit{Animal Battlegrounds}, \textit{Brainrot Animals}, \textit{Monky Bunch}, \textit{Monke Manic}, \textit{TOSS!}, \textit{Penguiin Paradise}, \textit{Raccoon Party}, \textit{Exploding Kittens}, \textit{Frog Roulette}, \textit{Fear Massive}, \textit{Five Nights At Monkeys}, \textit{GetaWay}, \textit{Gorilla Tag}, \textit{Boom Boom Hamster}, \textit{Big Monkey}, \textit{MonkeyWorks}, \textit{Ape-pocalypse}, \textit{Gorilla Football}, \textit{Raccoon Rampage}, \textit{UG}, \textit{Steal a Monke}, \textit{Penguin Festival} \\
        \bottomrule
    \end{tabularx}
\end{table}

\subsection*{Theme 1: Animal Biomimicry}

This theme encompasses design approaches that prioritize authentic reproduction of animal movement patterns and bodily behaviors. These games attempt to create an embodied experience that closely approximates how animals move and interact in their natural environments.

\subsubsection*{Specific Movement Pattern Simulation}

Games under this theme require players to perform actions that simulate animals' natural movements and bodily behaviors. For example, \textit{VR Pigeons} requires players to mimic the head-bobbing motion of walking pigeons to control locomotion, while \textit{Squid Smack} requires players to simulate the downward slapping motion of squid tentacles against water currents to move along the ocean floor.

\begin{figure}[h]
    \centering
    \includegraphics[width=0.6\linewidth]{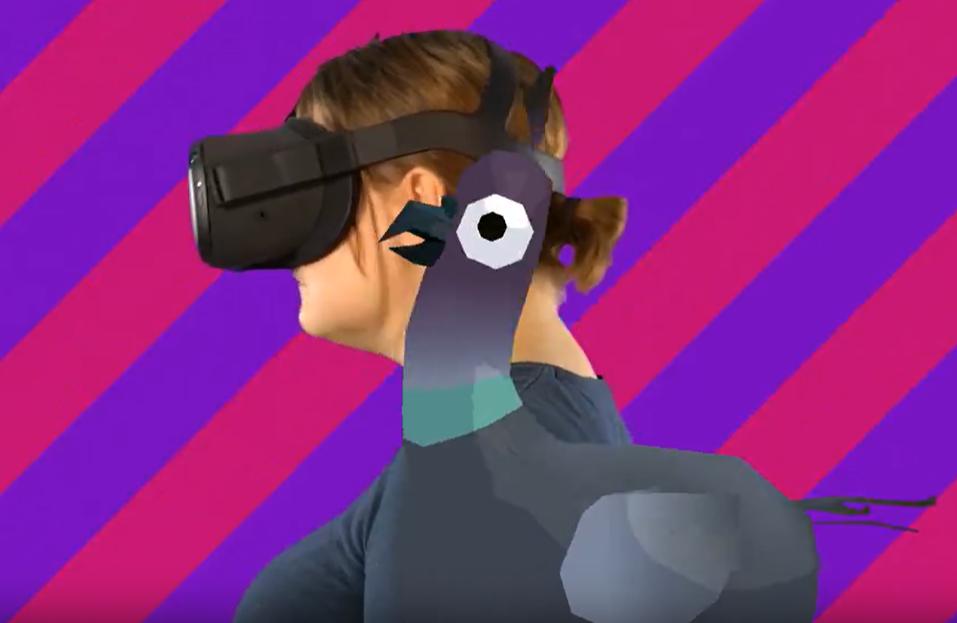}
    \caption{\textit{VR Pigeons}}
    \label{1}
\end{figure}

\subsubsection*{Coordinated Movement Beyond Mapped Points}

The head-mounted display and dual controllers serve as mapped tracking points in VR. Consequently, most VR games only model the upper body of animal avatars, specifically the head and hands, to meet in-game action requirements while reducing animal mimicry. Real-world animals do not move solely with their upper limbs, and most animals lack dexterous hands. In Theme 1, human players perform coordinated movements beyond the mapped points. In \textit{VR Pigeons}, players' head-bobbing motions to mimic pigeons drive movement in other body parts. Through this head-bobbing mechanism, players' full-body movements inadvertently resemble a pigeon walking and feeding on the ground, achieving action fidelity that would be difficult to replicate through hand gestures or controllers alone. In contrast, Theme 2 games primarily rely on interaction through the dual controller points. This distinction serves as the key differentiator between Themes 1 and 2.

\section*{Theme 2: Limited Animal Simulation}

This theme focuses on games that attempt to convey animal perception and bodily experience but do not transcend the limitations of current VR hardware and human physical constraints through their game mechanics.

\subsubsection*{Animal Perspective, Body Shape, and Scale Reproduction}

Games emphasizing visual and spatial authenticity present the world from an animal's perspective, including accurate body proportions, eye position, field of view, and environmental scale relative to the animal's body size. These VR games resemble the Simulator category in traditional 3D games, such as \textit{I Am Hamster - Simulator} and \textit{I Am Cat}, which primarily simulate the animal's view of the world through visual effects.

\begin{figure}[h]
    \centering
    \includegraphics[width=0.6\linewidth]{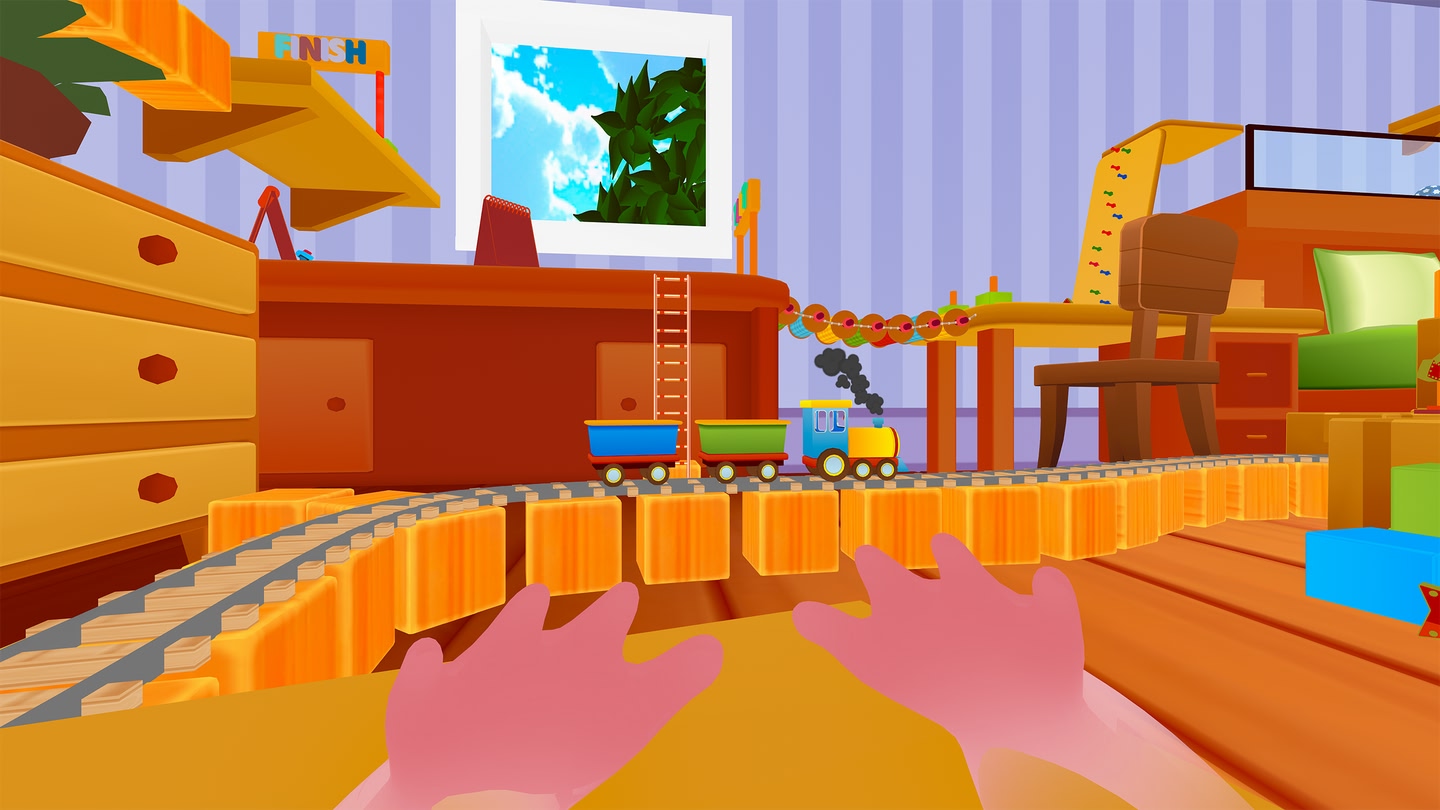}
    \caption{\textit{I Am Hamster - Simulator}}
    \label{2}
\end{figure}

\subsubsection*{Hardware-Constrained Movement}

Movement and interaction patterns are constrained or limited by VR hardware capabilities, including tracking limitations, controller configurations, and play space boundaries, resulting in simplified animal movements. Games under this theme continue to rely on controller-based locomotion. In \textit{Zombie Ants VR: Definitive Edition}, although players experience an ant's perspective, they still move and interact using controller-based raycast mechanics. In \textit{I Am Hamster - Simulator}, while the controllers are mapped to the hamster's paws with an avatar representation, interactions such as crawling and climbing lack distinctive characteristics. In other words, these interactions would function equally well for other animals.

\begin{figure}[h]
    \centering
    \includegraphics[width=0.6\linewidth]{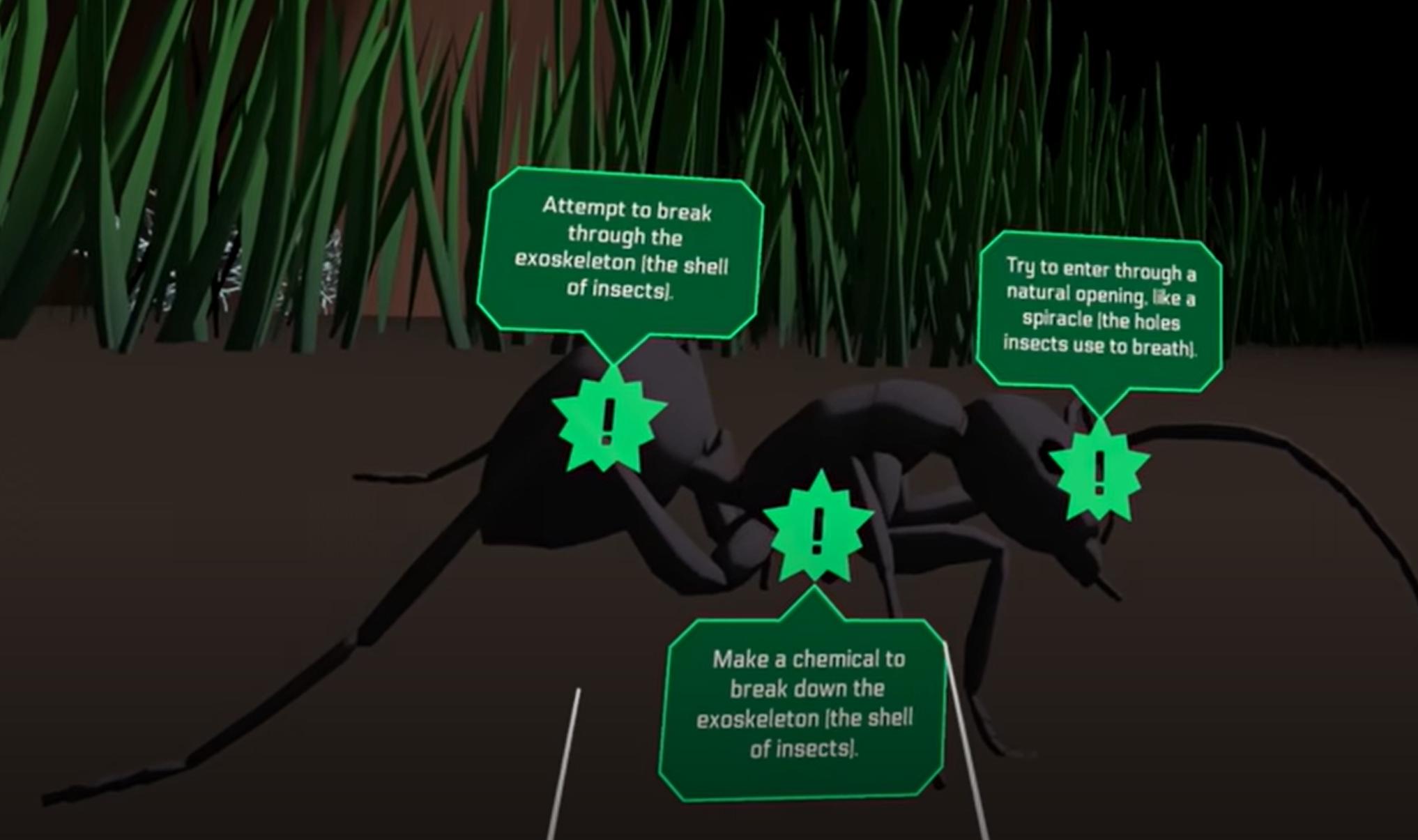}
    \caption{\textit{Zombie Ants VR: Definitive Edition}}
    \label{3}
\end{figure}

\section*{Theme 3: Hybrid Human-Animal Features}

This theme encompasses designs that merge human capabilities with animal characteristics.

\subsubsection*{Interactions Incorporating Animal Ecological Features}

This category includes design elements that utilize animal characteristics as interaction mechanisms, transforming species-specific abilities into playable game mechanics. In \textit{Tentacular}, players physically reach out and ``pop'' objects onto their tentacle. When players swing their arms, the tentacle follows with delay and weight.

\subsubsection*{Anthropomorphized Tasks}

To enhance interest in gameplay, some games add tasks that only humans can perform, such as tool usage and precision operation. In \textit{Tentacular}, players are asked to stack crates to build towers and use octopus tentacles to pull back giant rubber bands to launch objects or catch rockets. Later in the game, players use magnets to fuse junk together to create machines. These tasks are clearly executable only by humans. In \textit{Wreckin' Raccoon}, players enter a human kitchen from a raccoon's perspective to cause mischief. Although the perspective, body shape, and scale are authentic, and the game incorporates raccoon-specific features such as climbing, jumping, and burrowing, players can use kitchen utensils to cook, which clearly departs from the animal's natural attributes.

\begin{figure}[h]
    \centering
    \begin{minipage}[t]{0.45\linewidth}
        \centering
        \includegraphics[width=1\linewidth]{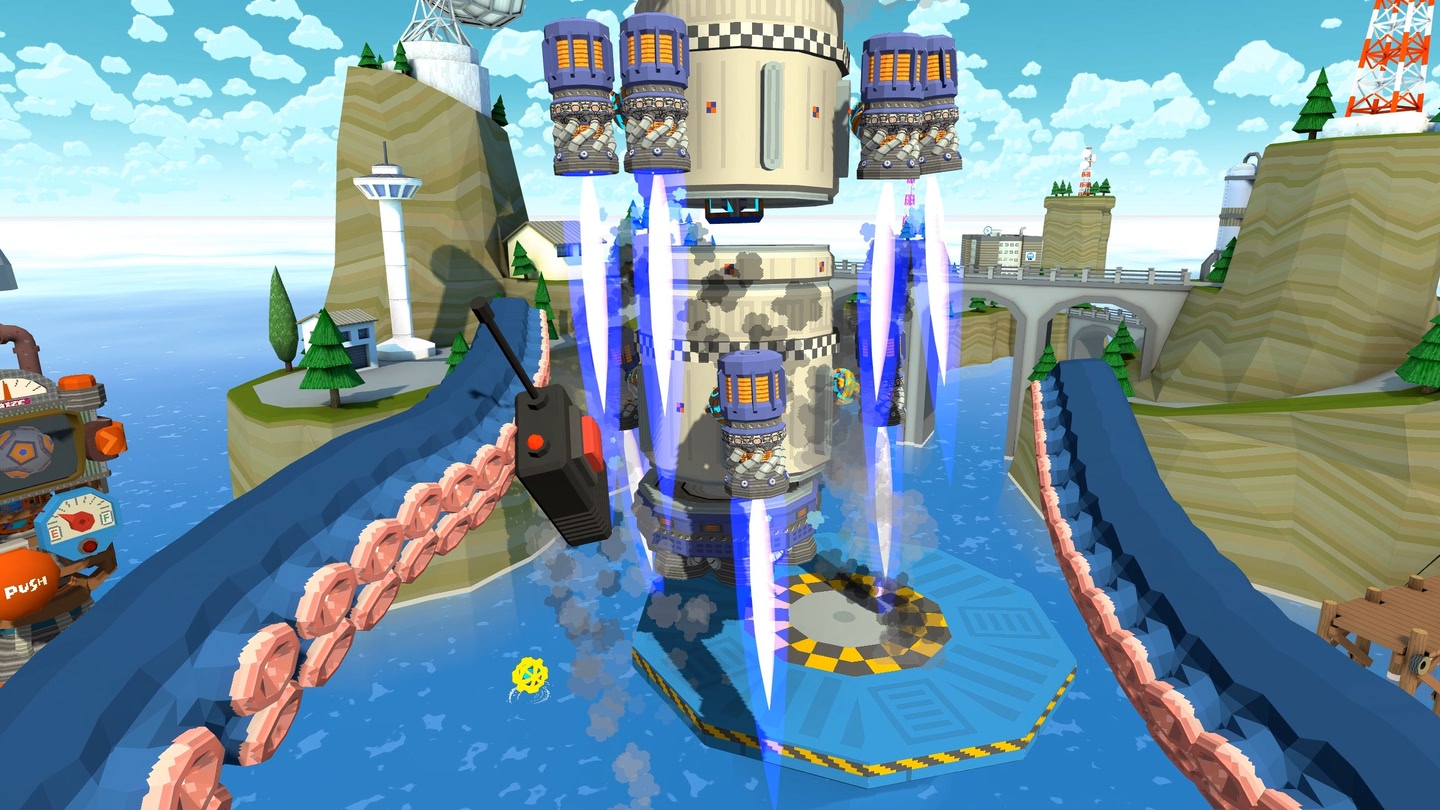}
        \caption{\textit{Tentacular}}
        \label{4}
    \end{minipage}
    \hfill
    \begin{minipage}[t]{0.45\linewidth}
        \centering
        \includegraphics[width=1\linewidth]{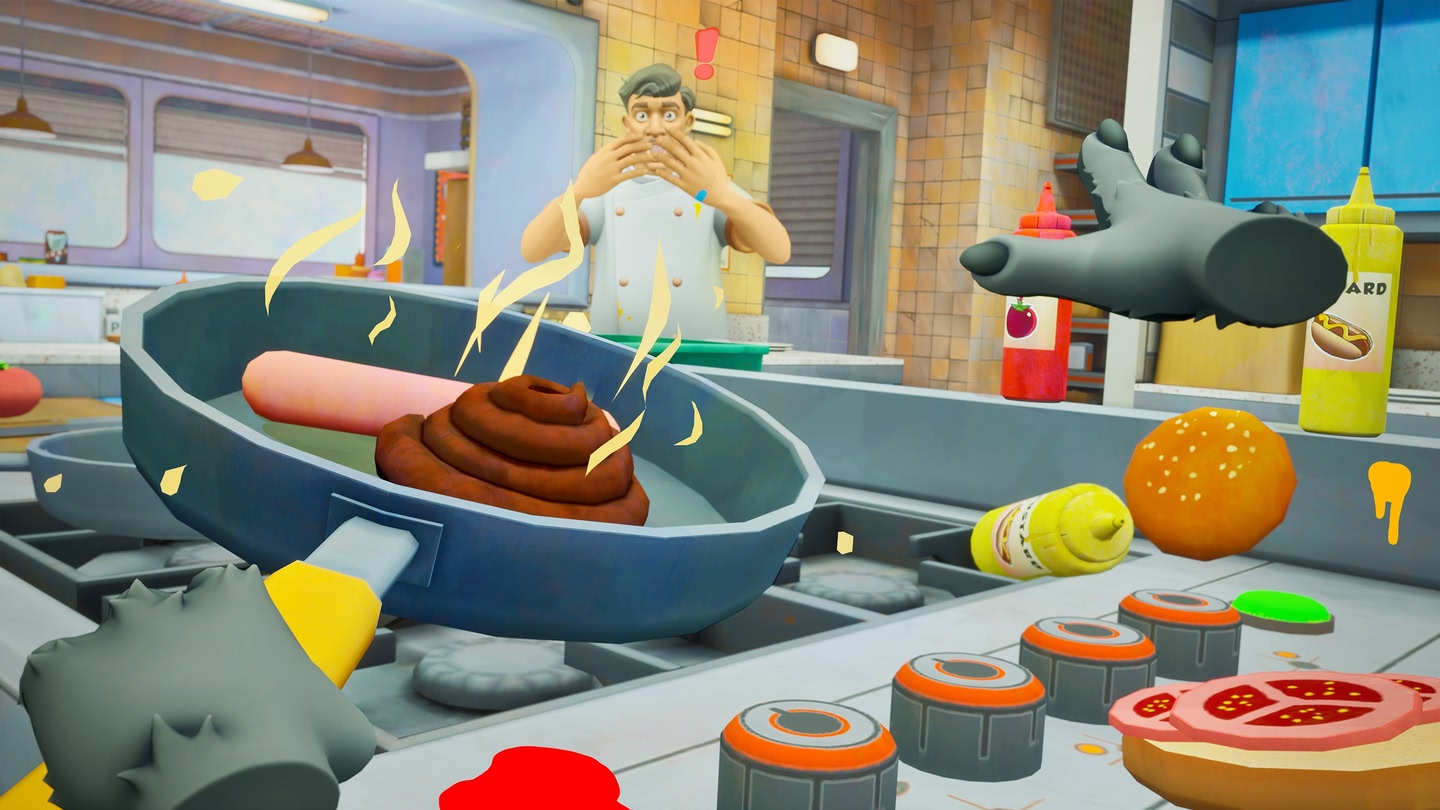}
        \caption{\textit{Wreckin' Raccoon}}
        \label{5}
    \end{minipage}
\end{figure}

\subsection*{Theme 4: Human Behavior with Animal Avatar}

This theme represents designs where human movement patterns and interaction logic remain dominant despite the use of animal avatar appearances. The animal form exists primarily as a visual shell rather than constituting a fundamentally different embodied experience.

\subsubsection*{Exclusively Human Movement and Interaction Logic}

Players interact using entirely traditional, human-centric control and movement patterns, including walking, grasping, and pointing, with the animal avatar serving only as visual representation. In \textit{Penguin Festival}, players controlling penguins can write on blackboards, bathe in hot springs, fish, and ski. In \textit{Boom Boom Hamster Doom}, players controlling hamsters can fire rockets at each other in intense combat. Beyond seeing animal models, players in these games cannot associate their experience with being embodied in an animal body.

\subsubsection*{Human Body as Foundation}
\begin{figure}[h]
    \centering
    \begin{minipage}[t]{0.45\linewidth}
        \centering
        \includegraphics[width=1\linewidth]{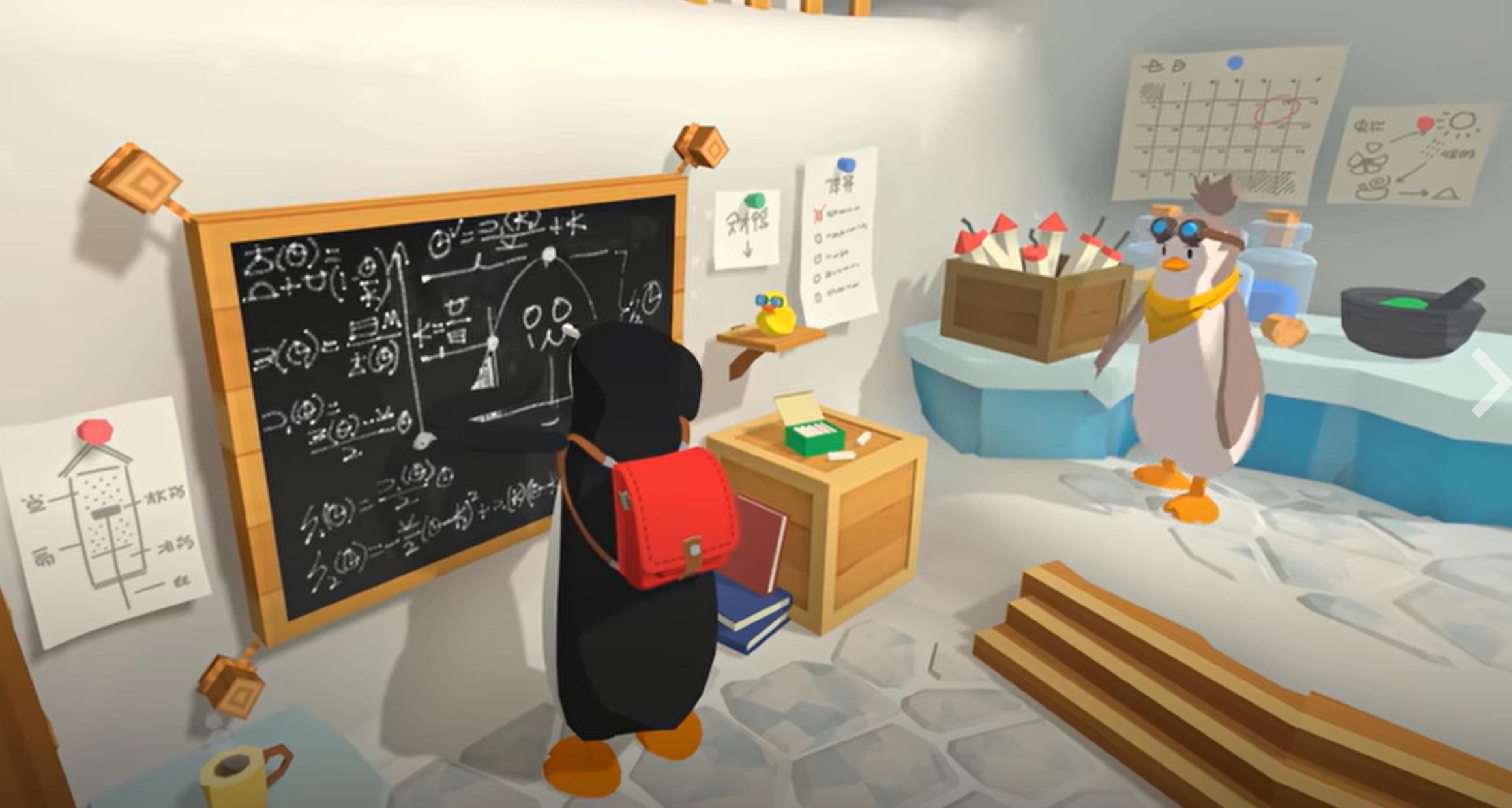}
        \caption{\textit{Penguin Festival}}
        \label{6}
    \end{minipage}
    \hfill
    \begin{minipage}[t]{0.45\linewidth}
        \centering
        \includegraphics[width=1\linewidth]{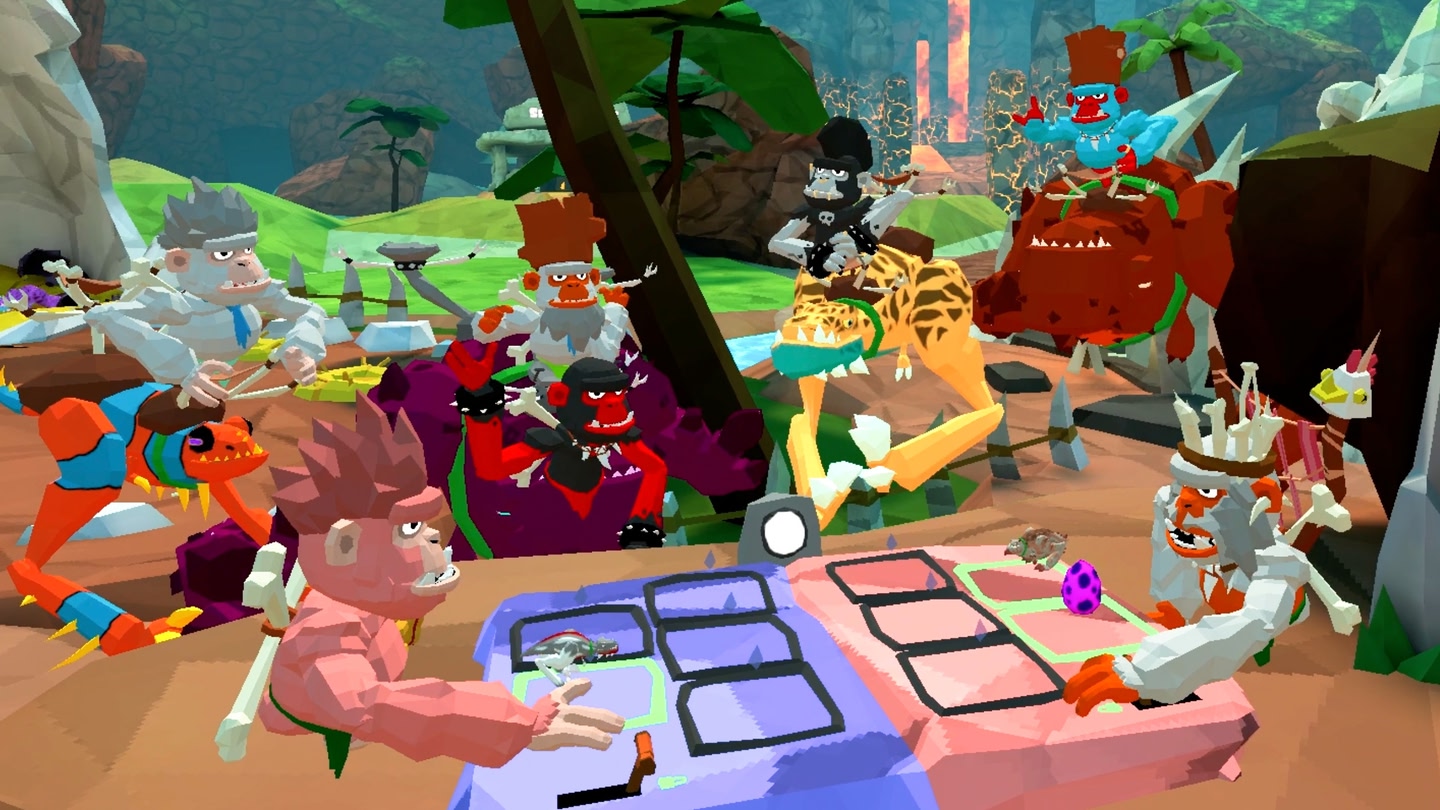}
        \caption{\textit{UG}}
        \label{7}
    \end{minipage}
\end{figure}

Avatar designs use human skeletal structure and proportions as the underlying framework, with animal features added only superficially in the form of anthropomorphized animals or animals with human proportions. In \textit{Penguin Festival}, the penguin avatar's skeletal rigging is nearly identical to human anatomy, making movement indistinguishable from human movement. The game perspective is not adjusted for actual penguin vision, and environmental proportions remain nearly identical to human scale. In \textit{UG}, players can embody various animals including dinosaurs, monkeys, and tigers, but only humanoid heads and hands are modeled, with no lower body representation.

\section*{DISCUSSION}
\begin{figure}[htbp]
    \centering
    \includegraphics[width=1\linewidth]{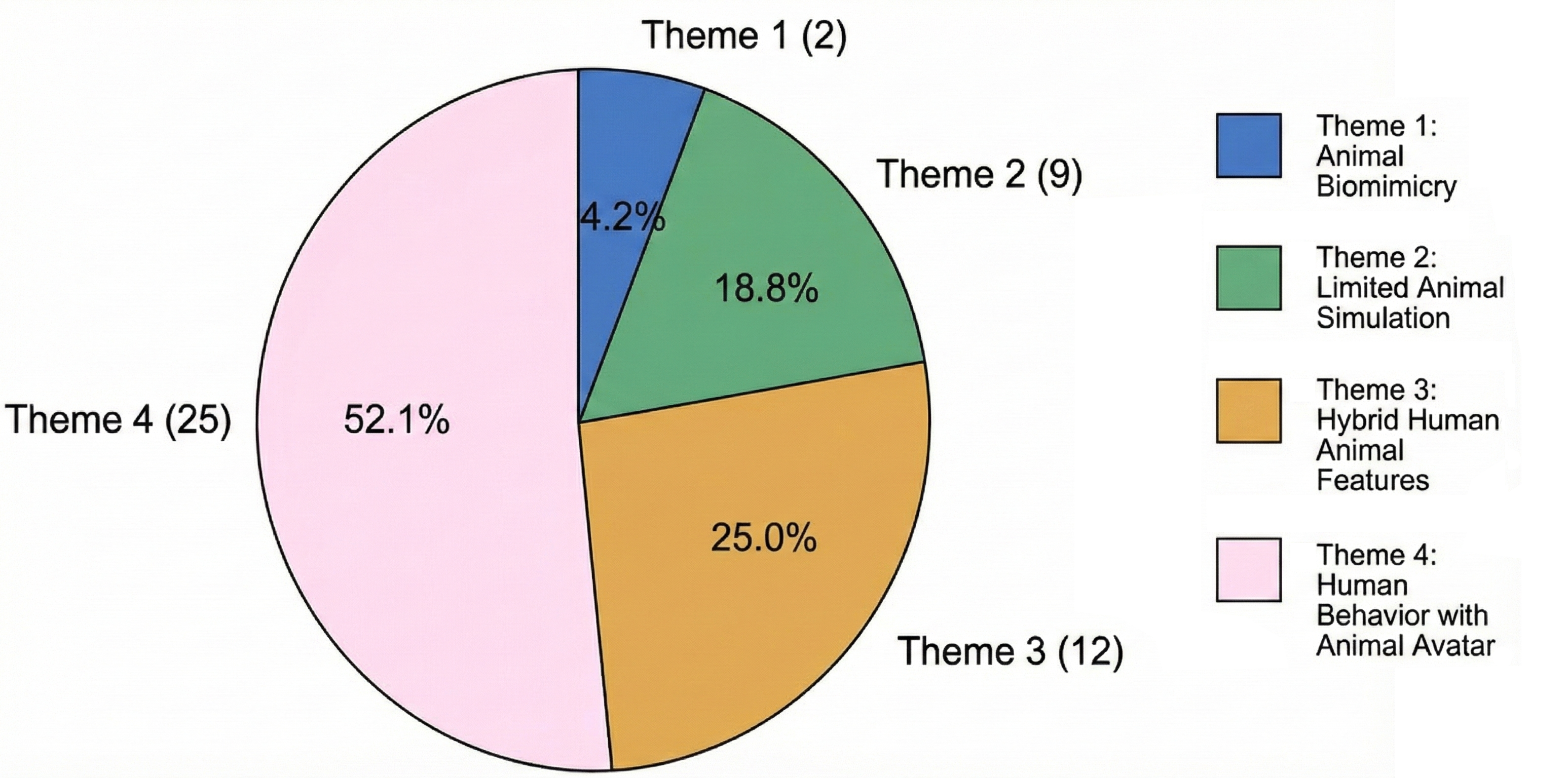}
    \caption{48 games classification results}
    \label{fig:placeholder}
\end{figure}

\subsection*{Distribution of Game Themes}

We categorized the 48 games according to the four themes identified in the upper part of paper, and the classification results are shown in the pie chart above (Figure 9). The results indicate: Theme 1 contains 2 games, accounting for approximately 4\%; Theme 2 has 9 games, accounting for approximately 19\%; Theme 3 has 12 games, accounting for approximately 25\%; Theme 4 has 27 games, accounting for approximately 56\%, making it the largest category. From the samples surveyed, approximately 77\% (Theme 1 + Theme 2) of the games are still fundamentally human-character-based games with an animal appearance layered on top, where the core gameplay and operational logic are essentially created from a human-centric perspective.

\subsection*{Problems in Game Design}

\subsubsection*{Human-Centric Design as the Origin Point}

After comprehensively reviewing games that feature first-person animal embodiment in VR, the problems now exposed are as mentioned above. Although these VR games provide players with animal avatars, the interaction logic in most games still follows human operation methods, lacking the capacity to express non-human body motion. For example, current VR controllers and tracking systems are predominantly designed based on human skeletal structure \cite{krekhov2019illusion}, which largely limits the possibility for players to experience authentic animal movement. The hardware and mindset starting from anthropocentrism means that existing VR controllers are designed based on human hand structure, causing developers to easily fall into Functional Fixedness. When players embody cats or birds, they still need to use buttons or triggers to interact, and this operational logic severely disconnects from the visual animal form, creating proprioceptive dissonance that destroys the illusion of becoming an animal. This is because existing human-centric VR input devices and control strategies cannot adequately express the body dynamics of non-human animals, preventing players from obtaining a true embodied experience. Furthermore, this lack of embodied experience contradicts the novel experience that the gaming medium emphasizes. Games typically use the promise of fresh, immersive experiences as a selling point, but if interactions are still designed according to human habits, it becomes difficult to satisfy players' expectations of truly becoming animals.

\subsubsection*{The Antinomy Between Authenticity and Comfort}

On the other hand, there is a contradiction between authenticity and comfort in VR animal embodiment game experiences, leading to a Design Antinomy situation. From design and philosophical perspectives, completely abandoning anthropocentrism is not necessary, the goal is not to pursue de-anthropocentrization for its own sake, but rather to utilize games as a rich multimedia platform for expressing meaning, philosophy, and conceptual exploration. Beyond the creator's perspective, novel embodied experiences provide players with fundamentally new modes of engagement. For example, human-centric design solutions provide players with familiar and comfortable operational logic, which is important for reducing learning costs and improving playability; however, if animal authenticity is overemphasized in design, it may sacrifice system controllability and entertainment value. For instance, having players on all fours like a tiger can indeed bring strong embodied sensation, but it leads to severe physical fatigue and motion sickness, and is constrained by domestic physical space. Conversely, if players are kept seated in chairs pressing buttons for comfort, they are turned back into drivers, losing the uniqueness of the animal experience. The current market mostly chooses to sacrifice authenticity in exchange for comfort and social entertainment, but this obscures VR's deeper potential as a transformative medium. Therefore, while ensuring basic usability, designers need to seek balance between the two. This exploration has important significance: by weighing human comfort against animal authenticity, new interactive experience spaces can be developed while providing possibilities for diversified VR gameplay.

\subsubsection*{The Significance of Absolute Simulation}

When examining substantive meaning, a theoretical question emerges: whether absolute animal simulation would increase or decrease player enjoyment, and what its significance is. Among the 48 surveyed games, socially-oriented games occupy a dominant market position, and from sales data, this indicates that players may not need to fully experience becoming animals in VR games to achieve satisfaction. This raises a key question: does the market's clear preference for social-function-dominant games represent genuine player needs, or is it a constrained choice determined by limited alternative options? Although we will conduct more in-depth qualitative analysis of this phenomenon in future research, the current landscape clearly lacks diverse methodological forms of animal embodiment.

Regarding possibilities for new forms, what significance do games pursuing absolute simulation have? Bidirectional analysis yields contrasting perspectives. On positive aspects: Collective animal role-playing (for example, everyone becoming dogs together and performing dog behaviors together) generates interesting social dynamics, as evidenced in animal-based games prevalent in early childhood education and team-building activities, such as Eagle Catches Chickens or What Time Is It, Mr. Wolf? Another side of negative aspects: Physical burden on the human body, poor sustainability, and spatial limitations present significant obstacles. These dimensions merit further empirical investigation and theoretical discussion.

\subsection*{Game Design Opportunities}

Although the market status quo tends toward conservatism, academia has already conducted some forward-looking research that provides ideas for resolving the aforementioned contradictions. Existing research indicates that non-human form virtual avatars have important development potential. Krekhov et al. found that non-human avatars with additional body parts can significantly enhance game enjoyment, and there is a strong correlation between virtual body ownership and game pleasure\cite{krekhov2019beyond}; Yu Jiang et al. emphasized the feasibility and superiority of using hand controllers to implement non-human avatars through proposing the HandAvatar method\cite{jiang2023handavatar}. Based on these perspectives, we believe that future non-human animal embodied interaction design can continue to develop in the following directions:

\begin{itemize}

\item \textbf{Controller Extensibility}: Design more flexible controllers to accommodate diverse animal body types. Krekhov et al. tested three animal prototypes—a rhino using full-body crawling with additional Vive trackers, a scorpion using upright half-body tracking, and a bird using arm-flapping for wing control—and found that while players achieved high game enjoyment and confirmed that additional body parts do not undermine IVBO, full-body tracking caused physical fatigue and standard controllers lacked physical correspondence with actual animal morphologies\cite{krekhov2019beyond}. Yu Jiang et al.'s HandAvatar method addresses part of this gap by enabling direct control of virtual animal joints through hand movements, demonstrating the potential of leveraging hand dexterity to drive non-humanoid bodies\cite{jiang2023handavatar}. The commercial games surveyed in this study reflect the same ceiling: games like \textit{Eagle Flight} and \textit{Tentacular} rely on standard hand controllers that map animal locomotion to button presses or arm gestures, creating proprioceptive dissonance between the player's physical input and the animal's movement logic. These works together suggest that future development could explore deformable or adaptive-morphology controllers—such as shape-changing handles or exoskeleton-like wearables—that physically align with the structures of diverse animal bodies, thereby strengthening proprioceptive feedback beyond what current tracker-dependent solutions afford.

\item \textbf{Non-Traditional Body Mapping}: Break away from one-to-one correspondence mapping patterns between human postures and animal forms. Krekhov et al. explored using IK solvers to map a single human leg to multiple animal limbs, and allowing an upright human posture to drive quadruped locomotion through half-body tracking—strategies that achieved IVBO comparable to or exceeding humanoid avatars while reducing physical strain\cite{krekhov2019illusion}\cite{krekhov2018vr}. However, participants noted that complex anatomical synchronization was sometimes imperfect, and finer-grained limb control remained elusive. Among the commercial games we surveyed, \textit{VR Pigeons} offers a notable counterexample: by requiring players to mimic the head-bobbing gait of a pigeon, full-body movement is coordinated through a single non-mapped tracking point, producing an emergent sense of animal locomotion that hand-controller games rarely achieve. Yet this approach is exceptional among the 48 games reviewed, with the vast majority defaulting to human-centric control logic. Future design should build on these experiments by developing AI-assisted hybrid mappings—for instance, dynamically translating arm gestures into whale flipper undulations or bird wing strokes—to enable more naturalistic animal movement while reducing cognitive load.

\item \textbf{Natural Dynamic Feedback}: Introducing dynamic feedback for animal-characteristic limbs can significantly enhance the embodied experience. Krekhov et al. demonstrated that attaching a synchronized virtual tail to a character significantly improves virtual body ownership, suggesting that animal-specific appendages can function as powerful embodiment enhancers rather than obstacles\cite{krekhov2019illusion}. Egeberg et al. further explored this by testing sensory feedback conditions for virtual wings, finding that visuomotor feedback was more effective than visuotactile feedback in strengthening agency and ownership over non-human appendages\cite{egeberg2016extending}. More recent work has extended this direction with richer modalities: Vargas et al. conducted two experiments on dog avatar embodiment and found that tactile feedback—real brushing stimulation synchronized with a virtual character's actions on the dog avatar—significantly improved body ownership and body change perception compared to baseline, while audio feedback, whether presented alone or combined with tactile feedback, produced no significant additive effect\cite{vargas2023dog}.Wang and Gao developed an avian avatar system using retractable straps to simulate wing mechanics and inflatable cushions to convey air resistance during flight, with results showing that spatial haptics augmented realistic wing sensations for the majority of participants, though highly personalized responses remained a limitation\cite{wang2024avian}. Future design could advance beyond this by incorporating dedicated haptic devices—such as vibrating wearable tails, pneumatic wing structures, or texture-conveying gloves—to deliver the kind of dynamic, animal-characteristic sensory feedback that current commercial implementations only approximate. 

\item \textbf{Cross-Media Experience}: Extend VR animal embodiment experiences to domains beyond gaming, enhancing educational and emotional resonance value. Delgado et al. developed a VR project in a museum setting that allowed children to swim with virtual marine creatures while receiving AI-guided educational narration, thus increasing learning interest and empathy for animals\cite{delgado2024enhancing}. Pimentel and Kalyanaraman showed that embodying a sea turtle in VR measurably increased participants' conservation-related behaviors, suggesting that embodiment grounded in animal-specific ecological context can produce real attitudinal change\cite{pimentel2022effects}. These cases demonstrate that the key lies in enabling users to understand animal behavioral logic from the animals' survival circumstances. In the future, this approach can be adopted to apply VR animal interaction to fields such as education, science communication, or art exhibitions, allowing different audiences to gain deeper understanding of themes such as animal behavior and environmental protection through immersive experiences.

\end{itemize}

In summary, building on existing research results, future embodied design for VR animal games can expand along directions such as controller diversification, innovative mapping, dynamic feedback, and cross-scenario applications to explore richer and more authentic non-human perspective interaction methods, thereby promoting innovation and development in VR interaction design.

\section*{CONCLUSION}
This study systematically analyzed 48 VR animal avatar games and proposed a design framework consisting of four themes to describe the main trends in current commercial design. The results show that most works still adhere to human-centered interaction logic. This contrasts with the theoretical potential of VR technology to support more radical bodily transformation. The gap between technological potential and practical application indicates that the field remains constrained by existing hardware paradigms, development conventions, and commercial risk considerations.

The theoretical contribution of this study lies in systematically mapping the design spectrum of VR animal avatars, ranging from animal biomimicry to fully human-like behavior. The framework highlights that the core tension in current design is the trade-off between authenticity and usability. This tension often stems not from technical limitations but from comprehensive considerations of player habits, device affordability, and product stability. At the same time, the study identifies several promising yet underdeveloped design directions, such as cross-morphology body mapping, natural dynamic feedback, and the introduction of novel interaction loci. These directions have been validated academically, but their commercial application remains in early stages.

This study has limitations in terms of samples and methodology. The samples were primarily drawn from mainstream platforms, making it difficult to cover niche or regional works; the thematic framework is based on the research team’s interpretation and requires further validation with more data; furthermore, the study focuses on the design level and does not explore how players perceive and respond to different degrees of animal authenticity. Future research could expand to the player level by collecting community comments, discussions, and behavioral data to build a richer dataset and further refine the boundaries and applicability of the four-theme classification framework.

From a cultural and media perspective, VR animal avatars involve not only technical implementation but also how users imagine and experience non-human existence. The highly anthropocentric nature of current design is not inevitable but rather the result of the interplay between technology, market, and culture. By revealing this structural tendency, this study lays the groundwork for advancing VR as a medium for perspective-shifting and embodied experience. Future work is expected to find more strategic balances between usability and animal authenticity, thereby deepening animal embodiment experiences from mere visual imitation to genuine embodied transformation.

\clearpage
\section*{BIBLIOGRAPHY}
\printbibliography

\clearpage
\appendix

\section*{APPENDIX}
\begin{figure}[h]
    \centering
    \includegraphics[width=1\linewidth]{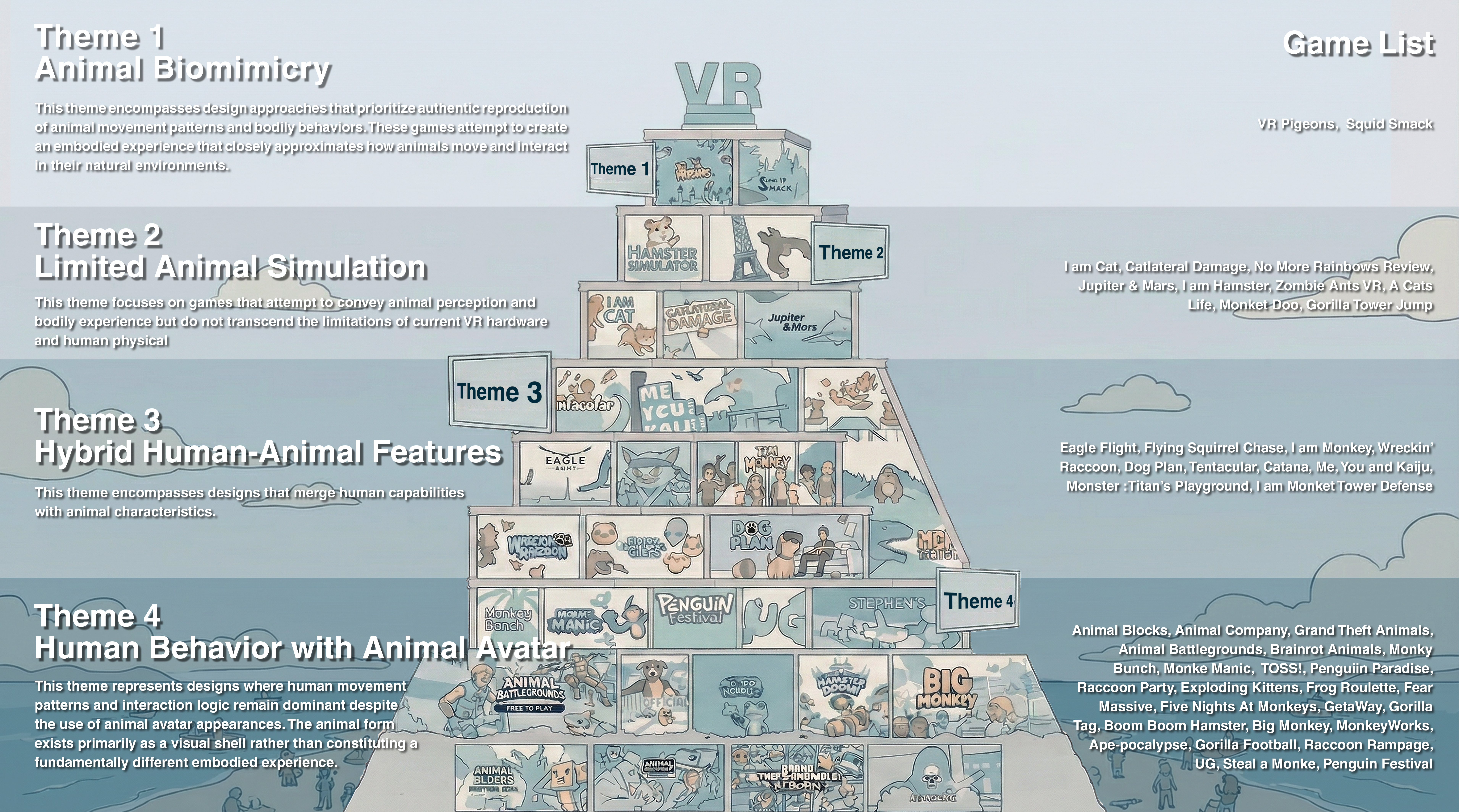}
    \caption{Four themes of animal embodiment in VR Games}
    \label{9}
\end{figure}

\end{document}